\documentstyle[preprint,floats,prl,aps,epsf]{revtex}

\begin{document}

\preprint{\hbox {April 2003} }

\draft
\title{Stability Issues for $w < -1$ Dark Energy}
\author{\bf Paul H. Frampton
}
\address{
Department of Physics and Astronomy,
University of North Carolina, Chapel Hill, NC  27599.}
\maketitle
\date{\today}

\begin{abstract}
Precision cosmological data
hint that a
dark energy with equation of state $w = P/\rho < -1$
and hence dubious stability is viable.
Here we discuss for any $w$ nucleation 
from $\Lambda > 0$ to $\Lambda = 0$ in a
first-order phase transition.
The critical radius is argued to be at least of galactic size
and the corresponding nucleation rate glacial,
thus underwriting the dark energy's stability and rendering
remote any microscopic effect.
\end{abstract}
\pacs{}

\newpage

\bigskip
\bigskip

\noindent {\it Introduction}.

\bigskip
\bigskip

\noindent The equation of state for the dark energy component
in cosmology has been the subject of much recent 
discussion
\cite{Melchiorri,Schuecker,Hannestad,Caldwell,Frampton,Bastero,Dicus,Ta,CHT}
Present data are consistent
with a constant  $w(Z) = -1$ corresponding to
a cosmological constant. But the data allow a present value
for
$w(Z=0)$ in the range
$- 1.38 < w(Z=0) < -0.82$ \cite{Melchiorri}.
If one assumes, more generally, that $w(Z)$ depends on $Z$
then the allowed range for $w(Z=0)$ is approximately the
same\cite{Ta}. In the present article we shall
forgo this greater generality as not relevant.
Instead, in the present article we address the question
of stability for a dark energy with constant 
$w(Z) < -1$.

\bigskip
\bigskip

\noindent {\it Interpretation as a limiting velocity}

\bigskip
\bigskip

\noindent Consider making a Lorentz boost along the
1-direction with velocity $V$ (put c = 1). Then
the stress-energy tensor which in the dark energy
rest frame has the form:
\begin{equation}
T_{\mu\nu} = \Lambda 
\left( 
\begin{array}{cccc}
1 & 0 & 0 & 0 \\
0 & w & 0 & 0 \\
0 & 0 & w & 0 \\
0 & 0 & 0 & w 
\end{array}
\right)
\label{restframe}
\end{equation}
is boosted to $T^{'}_{\mu\nu}$ given by

\begin{equation}
T^{'}_{\mu\nu} =
\Lambda
\left( 
\begin{array}{cccc}
1 & V & 0 & 0 \\
V & 1 & 0 & 0 \\
0 & 0 & 1 & 0 \\
0 & 0 & 0 & 1 
\end{array}
\right)
\left( 
\begin{array}{cccc}
1 & 0 & 0 & 0 \\
0 & w & 0 & 0 \\
0 & 0 & w & 0 \\
0 & 0 & 0 & w 
\end{array}
\right)
\left( 
\begin{array}{cccc}
1 & V & 0 & 0 \\
V & 1 & 0 & 0 \\
0 & 0 & 1 & 0 \\
0 & 0 & 0 & 1 
\end{array}
\right)
=
\Lambda
\left( 
\begin{array}{cccc}
1+V^2w & V(1+w) & 0 & 0 \\
V(1+w) & V^2+w & 0 & 0 \\
0 & 0 & w & 0 \\
0 & 0 & 0 & w 
\end{array}
\right)
\label{boost}
\end{equation}

\noindent We learn several things by studying Eq.(\ref{boost}).
First, consider the energy component $T_{00}^{'} = 1 + V^2w$.
Since $V < 1$ we see that for $w > -1$ this is positive $T_{00}^{'} > 0$
and the Weak Energy Condition (WEC) is respected\cite{hawking}.
For $w=-1$, $T_{00}^{'} \rightarrow 0$ as $V \rightarrow 1$ and
is still never negative. For $w< -1$, however, we see that 
$T_{00}^{'} < 0$ if $V^2 > -(1/w)$ and this violates the WEC
and is the first sign that the case $w < -1$ must be studied
with great care.
Looking at the pressure component $T_{11}^{'}$ we see the special
role of the case $w = -1$ because 
$w = T^{'}_{11}/T^{'}_{00}$ remains Lorentz invariant
as expected for a cosmological constant. Similarly
the off-diagonal components $T_{01}^{'}$
remain vanishing only in this case.
The main concern is the negativity of $T^{'}_{00} < 0$
which appears for $V^2 > -(1/w)$.
One possibility is that it is impossible
for $V^2 > -(1/w)$. The highest velocities known are
those for the highest-energy cosmic
rays which are protons with energy $\sim 10^{20}eV$.
These have $\gamma = (1 - V^2)^{-1/2} \sim 10^{11}$
corresponding to $V \sim 1 - 10^{-22}$. This would imply
that:
\begin{equation}
w > -1 - 10^{-22}
\label{sillylimit}
\end{equation}
which is one possible conclusion.

\bigskip
\bigskip

\noindent {\it First-Order Phase Transition and Nucleation Rate}.

\bigskip
\bigskip

\noindent But let us suppose, as hinted at by 
\cite{Melchiorri,Schuecker,Hannestad} 
that more precise cosmological data
reveals a dark matter which
violates Eq.(\ref{sillylimit}). Then, by boosting to an inertial
frame with $V^2 > -(1/w)$, one arrives at $T^{'}_{00} < 0$
and this would be a signal for vacuum instability\cite{PHF76}.
If the cosmological background
is a Friedmann-Robertson-Walker (FRW) metric the physics
is Lorentz invariant and so one should be able to see evidence
for the instability already in the preferred frame
where $T_{\mu\nu}$ is given by Eq.(\ref{restframe}).
This goes back to work in the 1960's and 1970's
where one compares the unstable vacuum to a superheated
liquid. As an example, at one atmospheric
pressure water can be heated carefully 
to above $100^0$ C without boiling.
The superheated water is metastable and attempts to nucleate
bubbles containing steam. However, there is an energy
balance for a three-dimensional bubble between the positive
surface energy $\sim R^2$ and
the negative latent heat energy of the 
interior $\sim R^3$ which
leads to a critical radius below which 
the bubble shrinks away and above which
the bubble expands and precipitates boiling\cite{Langer1,Langer2}.
For the vacuum the first idea in \cite{PHF76}
was to treat the spacetime vacuum as a 
four-dimensional material medium just like superheated water.
The second idea in the same paper 
was to notice that a hyperspherical bubble expanding at the speed of light
is the same to all inertial observers. This Lorentz invariance 
provided the mathematical  relationship
between the lifetime for unstable vacuum decay and
the critical radius of the four-dimensional 
bubble or instanton.

In the rest frame, the energy density is
\begin{equation}
T_{00} = \Lambda \sim (10^{-3} eV)^4 
\sim ({\rm 1 mm})^{-4}
\label{Lambda}
\end{equation}
since $10^{-3} eV \sim (1 {\rm mm})^{-1}$. 

In order to make an estimate of the 
dark energy decay lifetime
in the absence of a known potential,
we can proceed by assuming (without motivation from observation)
that
{\it there is a first-order phase transition possible between the 
$\Lambda = (10^{-3} eV)^4$ ``phase'' and a $\Lambda = 0$ ``phase''}.
This hypothesized decay is the Lorentz invariant 
process of a hyperspherical bubble
expanding at the speed of light, the same for all
inertial observers.
Let the radius of this hypersphere be R, its energy density be $\epsilon$
and its surface tension be $S_1$. Then according to \cite{PHF76}
the relevant instanton action is
\begin{equation}
A = -\frac{1}{2} \pi^2 R^4 \epsilon + 2 \pi^2 R^3 S_1
\label{action}
\end{equation}
where $\epsilon$ and $S_1$ are the volume and
surface energy densities, respectively.
The stationary value of this action is
\begin{equation}
A_m = \frac{27}{2} \pi^2 S_1^4 /\epsilon^3
\label{stationaryA}
\end{equation}
corresponding to the critical radius
\begin{equation}
R_m = 3S_1 / \epsilon
\label{Rcritical}
\end{equation}
We shall assume 
that the wall thickness is negligible compared to
the bubble radius.
The number of vacuum nucleations in the
past lightcone is estimated
as
\begin{equation}
N = (V_u \Delta^4) exp ( - A_m)
\label{nucleations}
\end{equation}
where $V_u$ is the 4-volume of the past and
$\Delta$ is the mass scale relevant to the
problem.
This vacuum decay picture led
to the proposals of inflation\cite{Guth},
for solving the horizon, flatness and monopole problems 
(only the horizon problem was generally known at
the time of \cite{PHF76}).
None of that work addressed why the true vacuum has zero energy.
Now that the observed vacuum has non-zero energy
density
$+ \epsilon \sim (10^{-3} eV)^4$
we may interpret it as a false vacuum lying
above the true vacuum with $\epsilon = 0$.
In order to use the full power 
of Eq.(\ref{nucleations})
taken from \cite{PHF76,PHF77} and the
requirement $N \ll 1$ we need
to estimate the three mass-dimension parameters
$\epsilon^{1/4}, S_1^{1/3}$ and $\Delta$ therein
and so we discuss these three scales in turn.

The easiest of the three to select is $\epsilon$. If we imagine a
tunneling through a barrier between a false vacuum
with energy density $\epsilon$ to a true vacuum at energy density zero
then the energy density inside the bubble will
be $\epsilon = \Lambda = (10^{-3} eV)^4$. No other
choice is reasonable.
As for the typical mass scale $\Delta$ in the prefactor
of Eq. (\ref{nucleations}), the value of $\Delta$
does not matter very much because it appears in a power
rather than an exponential so let us put (the reader 
can check that the
conclusions do not depend on this choice)
$\Delta = \epsilon^{1/4} = (1 mm)^{-1}$ whereupon the
prefactor in Eq.(\ref{nucleations}) is $\sim 10^{116}$.
The third and final scale to discuss 
is the surface tension, $S_1$.
Here we appeal to comparison of spontaneous decay to stimulated
decay. The former dictates $N \ll 1$ in 
Eq.(\ref{nucleations}): the latter requires further discussion.

Spontaneous dark energy decay brings us to the question of whether such
decay can be initiated in an environment existing
within our Universe. The question is analogous to
one of electroweak phase transition in 
high energy particle collision.
This was first raised in \cite{PHF76}
and revisited for cosmic-ray collisions
in \cite{Hut}. That was in the context of
the standard-model Higgs vacuum and the
conclusion is that high-energy colliders
are safe at all present and planned foreseeable energies because
much more severe conditions have already
occurred (without disaster) in cosmic-ray
collisions within our galaxy.
More recently, this issue has been addressed
in connection with fears
that the Relativistic Heavy Ion Collider
(RHIC) might initiate a diastrous 
transition
but according to careful analysis\cite{JBWS,AdR}
there was no such danger. 

The energy density involved for dark energy is
some 58 orders of magnitude smaller [$(10^{-3}eV)^4$
compared to $(300 GeV)^4$] than for the electroweak
case and so the nucleation scales are completely different.
One is here led away from microscopic towards 
astronomical size scales.

The energy density of Eq.(\ref{Lambda}) is so readily exceeded
that the critical radius cannot be
microsopic. Think first of a macroscopic scale {\it e.g. } 
1 meter and consider a magnetic field
practically-attainable in bulk on Earth 
such as 10 Tesla. Its energy density
is given by

\begin{equation}
\rho_{mag} = \frac{1}{\mu_0} {\cal B}^2
\label{magnet}
\end{equation}
Using the value $\mu_0 = 4 \pi \times 10^{-7} N A^{-2}$
and $1 T = 6.2 \times 10^{12} (MeV.s.m^{-2})$ leads
to an energy density
$\rho_{mag} = 2.5 \times 10^{17} eV/(mm)^3$, over 20 orders 
above the value of Eq.(\ref{Lambda}) for the interior
of  nucleation. Magnetic fields in bulk exist
in galaxies with strength $\sim 1 \mu G$ and the
rescaling by ${\cal B}^2$ then would
give
$\rho_{mag} \sim (2.5 \times 10^{-5} eV)(mm)^{-3}$,
slightly below the value in Eq.(\ref{Lambda}). {\it Assuming} the
dark energy
can exchange energy with magnetic energy density
the observed absence of stimulated decay would then 
imply a critical radius of at least
galactic size, say, $\sim 10 kpc.$
Using Eq(\ref{Rcritical}) then gives for the surface tension
$S_1 > 10^{23} (mm)^{-3}$ and number
of nucleations in Eq.(\ref{nucleations}) 
$N < exp (- 10^{92})$. The spontaneous decay is thus glacial.  
Note that the dark energy has appeared only recently
in cosmological time and has never interacted with
background radiation of comparable energy density.
Also, this nucleation argument does not require $w < -1$.
\bigskip
\bigskip

\noindent {\bf Discussion}

\bigskip
\bigskip

As a first remark, since the critical radius $R_m$
for nucleation is astronomical, it appears that
the instability cannot be triggered by any 
microscopic process.
While it may be comforting to know that the 
dark energy is not such a doomsday phenomenon, it
also implies at the same time the dreadful conclusion 
that dark energy may have no 
microscopic effect.
If any such microscopic effect in a terrestrial 
experiment could be 
found, it would be crucial in 
investigating the dark energy phenomenon.
We note that the present arguments
are less model-dependent than those in \cite{CHT}.

In closing one may speculate 
how such stability arguments may evolve.
One may expect most conservatively that
the value $w = -1$ will eventually be established empirically in which case 
both quintessence and the ``phantom menace''
will be irrelevant. 
In that case, indeed for any $w$, we may still hope that dark energy will
provide the first connection
between string theory and the real
world as in {\it e.g.} \cite{Bastero}.
Even if precise data do establish $w < -1$,
as in the ``phantom menace'' scenario, the 
dark energy stability issue is still under control.

\bigskip
\bigskip
\bigskip

\noindent {\it Acknowledgements}

\bigskip

We thank Stuart Raby and Tomo Takahashi for useful discussions.
This work
is supported in part by the
US Department of Energy under
Grant No. DE-FG02-97ER-41036.


\newpage




\begin{thebibliography}{99}
\bibitem{Melchiorri}
A. Melchiorri, L. Mersini, C.J. Odman and M. Trodden.
{\tt astro-ph/0211522}.
\bibitem{Schuecker}
P. Schuecker, R.R. Caldwell, H. Bohringer, C.A. Collins and L. Guzzo.
{\tt astro-ph/0211480}.
\bibitem{Hannestad}
S. Hannestad and E. M\"{o}rtsell.
Phys. Rev. {\bf D66,} 063508 (2002). {\tt astro-ph/0205096}.
\bibitem{Caldwell}
R.R. Caldwell,
Phys. Lett. {\bf B545,} 23 (2002).
{\tt astro-ph/9908168}.
\bibitem{Frampton}
P.H. Frampton, Phys. Lett. {\bf B555,} 139 (2003).
 {\tt astro-ph/0209037}.
\bibitem{Bastero}
M. Bastero-Gil, P.H. Frampton and L. Mersini, Phys. Rev. {\bf D65,}
106002 (2002).
\bibitem{Dicus}
D.A. Dicus and W.W. Repko.
{\tt hep-ph/0211109}.
\bibitem{Ta}
P.H. Frampton and T. Takahashi, Phys. Lett. {\bf B557,} 135 (2003).
{\tt astro-ph/0211544}.
\bibitem{CHT}
S.M. Carroll, M. Hoffman and M. Trodden. {\tt astro-ph/0301273}
\bibitem{hawking}
S.W. Hawking, Comm. Math. Phys. {\bf 25,} 152 (1972).\\
L.H. Ford. {\tt gr-qc/0301045}. 
\bibitem{Langer1}
J. Langer, Ann. Phys. {\bf 41,} 108 (1967).
\bibitem{Langer2}
J. Langer, Ann. Phys. {\bf 54,} 258 (1969).
\bibitem{PHF76}
P.H. Frampton, Phys. Rev. Lett. {\bf 37,} 1378 (1976).
\bibitem{Guth}
A.H. Guth, Phys. Rev. {\bf D23,} 347  (1981).\\
A.D. Linde, Phys. Lett. {\bf B108,} 389 (1982).\\
A. Albrecht and P.J. Steinhardt, Phys. Rev. Lett. {\bf 48,} 1220 (1982).
\bibitem{PHF77}
P.H. Frampton, Phys. Rev. {\bf D15,} 2922 (1977). 
\bibitem{Hut}
P. Hut, Nucl. Phys. {\bf A418,} 301C (1984).\\
P. Hut and M.J. Rees, IAS-Princeton preprint 83-0042 (1984)
\bibitem{JBWS}
R.L. Jaffe, W. Busza, F. Wilczek and J. Sandweiss,
Rev. Mod. Phys. {\bf 72,} 1125 (2000).
\bibitem{AdR}
A. Dar, A. De Rujula and U. Heinz, Phys. Lett. {\bf B470,} 142 (1999).

\end{thebibliography}
\end{document}